\documentclass[11pt]{article}
\usepackage{jheppub} 
\usepackage{slashed}
\usepackage{empheq}
\usepackage{blkarray}
\usepackage{tikz}
\usetikzlibrary{arrows,decorations.pathmorphing,backgrounds,positioning,fit,petri,automata,shadows,calendar,mindmap,
decorations.markings,calc}
\def\nn{\nonumber\\ }
\def\rd{{\rm d}}
\def\lsix{ \mathcal{L}^{(6)}}
\def\vev#1{\left\langle #1 \right\rangle}
\begin{document}

\title{Naive Dimensional Analysis Counting of Gauge Theory Amplitudes and Anomalous Dimensions}

\author{Elizabeth E.~Jenkins$^1$,}

\affiliation{1. Dept. Physics, University of California, San Diego, 9500 Gilman Drive,
  La Jolla, CA 92093-0319}

\author{Aneesh V.~Manohar$^1$,}


\author{Michael Trott$^2$}

\affiliation{2. Theory Division, Physics Department, CERN, CH-1211 Geneva 23, Switzerland}

\abstract{
We show that naive dimensional analysis (NDA) is equivalent to the result that $L$-loop scattering amplitudes 
have perturbative order $N=L+\Delta$,  with a shift $\Delta$ that depends on the NDA-weight of operator insertions. The NDA weight of an operator is defined in this paper, and the general NDA formula for perturbative order $N$  is derived.  The formula is used to explain why the one-loop anomalous dimension matrix for dimension-six operators in the Standard Model effective field theory has entries with perturbative order ranging from 0 to 4. The results in this paper are valid for an arbitrary effective field theory, and they constrain the coupling constant dependence of anomalous dimensions and scattering amplitudes in a general effective field theory.
}

\maketitle

\section{Introduction}

The LHC runs at $7$ and $8$ ${\rm TeV}$ have shown that the Standard Model (SM) with a single Higgs doublet $H$ provides a good description of the 
data up to an energy scale of a few ${\rm TeV}$.    
A reasonable interpretation of this experimental data is that the SM is a valid Effective Field Theory (EFT) up to an energy scale $\Lambda$ of order a few $ {\rm TeV}$, and that any new physics beyond the SM arises at energies $E \ge \Lambda$.   This interpretation also is supported by the measurements of rare decay processes in the complementary flavor physics program, which provides independent experimental evidence that the SM is a valid EFT up to energies of at least several $\rm TeV$.

The leading operators which parametrize arbitrary new physics effects in the SM EFT for energies $E \le \Lambda$ are of mass dimension six.
Recently, the one-loop anomalous dimension matrix $\gamma$ of the dimension-six operators in the SM was studied as part of an analysis of LHC results such as $h \to \gamma\gamma$ decay~\cite{Grojean:2013kd,Jenkins:2013zja}. 
Ref.~\cite{Jenkins:2013zja}  showed that $\gamma$ only has terms of order 
\begin{align}
\frac{1}{16\pi^2} \times \left\{1,g^2,\lambda,y^2,g^4,g^2\lambda,g^2y^2,\lambda^2,\lambda y^2,y^4,g^6,  g^4 \lambda, g^6\lambda \right\}\,,
\label{1}
\end{align}
where $g$ is a gauge coupling, $y$ is a Yukawa coupling, and $\lambda$ is the Higgs self-coupling.
Ref.~\cite{Jenkins:2013zja}  studied this pattern and showed that under a simple rescaling of the dimension-six operators based on naive dimensional analysis 
(NDA)~\cite{Manohar:1983md}, the one-loop anomalous dimension matrix elements have the form
\begin{align}
\left( \frac{g^2}{16\pi^2}\right)^{n_1} \left( \frac{\lambda}{16\pi^2}\right)^{n_2} \left( \frac{y^2}{16\pi^2}\right)^{n_3},\qquad
N=n_1+n_2+n_3
\label{2}
\end{align}
where $N \equiv n_1+n_2+n_3$, the perturbative order of the term, ranges from $N=0$ to $N=4$, even though all of the terms are from one-loop diagrams.

In this paper, we derive a formula for the perturbative order $N$ of a scattering amplitude at $L$ loops in an arbitrary EFT.  The result is equivalent to the NDA formula in Ref.~\cite{Manohar:1983md}. Our result, given in Eq.~(\ref{wt}), shows that $N$
depends on the  number of loops $L$, as well as on the NDA-weights of the operators, which we define below.
The proof is simple --- the NDA result of Ref.~\cite{Manohar:1983md} counts powers of $4\pi$ in the denominator. The NDA $4\pi$-counting formula then determines $N$, since the power of $4\pi$ in the denominator of
Eq.~(\ref{2}) is $2N$.  The formula derived is of significant value. As one application, it predicts the structure of anomalous dimension matrices
at any loop order in an arbitrary EFT.

\section{The NDA Counting Formula}

We derive the NDA-weight formula for a general EFT, i.e.\ a gauge theory with arbitrary higher dimension gauge invariant operators.
The EFT Lagrangian is given by
\begin{align}
{\cal L} &= {\cal L}^{(d \le 4)} + \sum_{d > 4} {\cal L}^{(d)}
\end{align}
where ${\cal L}^{(d \le 4)}$ denotes the terms in the effective Lagrangian with mass dimension $d \le 4$ and ${\cal L}^{(d)}$ denotes the terms in the effective Lagrangian with mass dimension $d$.
The general form of ${\cal L}^{(d \le 4)}$ is 
\begin{align}
\mathcal{L}^{(d \le 4)} &= -\frac14 X^2 + \overline \psi \, i \slashed{D} \psi + \frac12 (D \phi)^2 - M \overline \psi \psi -\frac12 m^2  \phi^2
-y \phi \overline \psi \psi - \kappa \phi^3 - \lambda \phi^4\,, 
\label{lag}
\end{align}
where gauge field strength tensors are denoted by $X_{\mu \nu}$, fermion fields by $\psi$, scalar fields by $\phi$, gauge couplings by $g$, fermion masses by $M$, scalar masses by $m$, Yukawa couplings by $y$, scalar cubic couplings by $\kappa$, and scalar quartic couplings by $\lambda$.  The covariant derivative is 
$D^\mu = \partial^\mu + i \sum_k g_k A^\mu_k$ where the sum on $k$ is over all gauge groups.
The $d \le 4$ Lagrangian can also contain topological terms, but these terms are not important for the present discussion and have been omitted.

The NDA formula of Ref.~\cite{Manohar:1983md} says that a generic term in an effective Lagrangian generated at a scale $\Lambda$ should be written as
\begin{align}
f^2 \Lambda^2 \left( \frac{\phi}{f} \right)^A \left( \frac{\psi}{f \sqrt \Lambda} \right)^B   \left( \frac{g X}{\Lambda^2} \right)^C 
 \left( \frac{D}{\Lambda} \right)^D\,,
 \label{nda}
\end{align}
with $\Lambda \sim 4\pi f$.  Using this normalization, Eq.~(\ref{lag}) can be written as
\begin{align}
\mathcal{L}^{(d \le 4)} &= -\frac1{4 \widehat C_X}  \frac{f^2 g^2 X^2}{\Lambda^2}  + \frac{1}{\widehat C_\psi} \overline \psi \, i \slashed{D} \psi + \frac{1}{2\widehat C_{\phi} } (D \phi)^2-\widehat C_M \Lambda \overline \psi \psi - \frac12 \widehat C_{m^2} \Lambda^2  \phi^2 \nn
&-\widehat C_y \frac{ \Lambda \phi \overline \psi \psi }{f} - \widehat C_\kappa \frac{\Lambda^2 \phi^3}{f} - \widehat C_\lambda \frac{\Lambda^2 \phi^4}{f^2}\,,
\label{rlag}
\end{align}
with rescaled inverse kinetic term coefficients and rescaled couplings given by
\begin{align}
\widehat C_X &= \frac{f^2 g^2}{\Lambda^2}=\frac{g^2}{16\pi^2}, & \widehat C_\psi &= 1,  & \widehat C_{\phi} &=1,  \nn
\widehat C_M &= \frac{M}{\Lambda},  & \widehat C_{m^2} &= \frac{m^2}{\Lambda^2},  
& \widehat C_y &= \frac{ y f}{\Lambda}=\frac{y}{4\pi}, \nn
\widehat C_\kappa &= \frac{\kappa f}{\Lambda^2} =\frac{1}{4\pi} \frac{\kappa}{\Lambda}, 
& \widehat C_\lambda &= \frac{\lambda f^2}{\Lambda^2} = \frac{\lambda}{16\pi^2},& &
\label{chat}
\end{align}
setting $f = \Lambda /4 \pi$.
The kinetic term coefficients are defined as $\widehat C^{-1}$, since the particle propagators, which enter the diagrams, are the inverse of the kinetic energy terms.  We can also define a second version of Eq.~(\ref{rlag}),  in which we do not distinguish between $f$ and $\Lambda$ by the substitutions $\widehat C_i \to C_i$,
with the $C_i$ evaluated for $f = \Lambda$
\begin{align}
C_X &=g^2, & C_\psi &=  1, & C_\phi &=1, \nn
C_M &= \frac{M}{\Lambda},  & C_{m^2} &=  \frac{m^2}{\Lambda^2},  & C_y &= y, \nn
C_{\kappa} &= \frac{\kappa}{\Lambda} , & C_\lambda &= \lambda\, . &&
\label{chat}
\end{align}
We call this normalization convention standard dimensional analysis.
In both cases, $\Lambda$ drops out of all dimension-four operator couplings.  The NDA coupling constants  $\widehat C_i$
differ from the coupling constants $C_i$ of standard dimensional analysis by  factors of
of $4\pi$.

Similarly, the operators in the effective Lagrangian ${\cal L}^{(d)}$ of dimension $d$ can be normalized according to NDA and standard dimensional analysis,\begin{align}
\mathcal{L}^{(d)} &= \sum \widehat C_{d,w} \frac{O_{d,w}}{f^{2w} \Lambda^{d-4-2w}} =  \sum  C_{d,w} \frac{O_{d,w}}{\Lambda^{d-4}} 
\label{ndaw}
\end{align}
where $w$, the power of $f^2$ in the denominator, is defined to be the NDA-weight of the operator, and the NDA coefficient are denoted by $\widehat C_{d,w}$. The standard dimensional analysis coefficients $C_{d,w}$ are for the same operator $O_{d,w}$ normalized only using powers of $\Lambda$ (i.e.\ with $f \to \Lambda$), but still including a factor of $g$ with each $X$.
As an example, the SM dimension-six operators fall into eight classes, 
\begin{align}
& \frac{g^3 X^3 }{\Lambda^2} && \frac{\phi^6}{\Lambda^2}, &&  \frac{\phi^4 D^2 }{\Lambda^2}, &&  \frac{g^2 X^2 \phi^2}{\Lambda^2} , 
&& \frac{ \psi^2 \phi^3}{\Lambda^2}, &&  \frac{\psi^2 g X \phi }{\Lambda^2} , && \frac{ \psi^2 \phi^2 D}{\Lambda^2}, 
&& \frac{ \psi^4} {\Lambda^2} \nonumber \\[5pt]
& \frac{f^2 g^3 X^3 }{\Lambda^4} , && \frac{\Lambda^2 \phi^6}{f^4}, &&  \frac{\phi^4 D^2 }{f^2}&&  \frac{g^2 X^2 \phi^2}{\Lambda^2} , &&
\frac{\Lambda \psi^2 \phi^3}{f^3}, &&  \frac{\psi^2 g X \phi }{\Lambda f} , && \frac{ \psi^2 \phi^2 D}{f^2}, && \frac{ \psi^4} {f^2} 
\label{ops}
\end{align}
with NDA-weights $w=\left\{-1,2,1,0,3/2,1/2,1,1 \right\}$, respectively. The upper row is the usual $C_{d,w}$ normalization, the lower row is the NDA $\widehat C_{d,w}$ normalization, and Eq.~(\ref{ndaw}) is
\begin{align}
\mathcal{L}^{(6)} &= C_{6,-1} \frac{g^3 X^3}{\Lambda^2} + \ldots  = \widehat C_{6,-1} \frac{f^2 g^3 X^3}{\Lambda^4} + \ldots\,.
\end{align}

The NDA result of Ref.~\cite{Manohar:1983md} is that an arbitrary graph in the EFT generates a scattering amplitude $\widehat C_{d,w}$ corresponding to $O_{d,w}/f^{2w} \Lambda^{d-4-2w}$ with
\begin{align}
\widehat C_{d,w} &\propto  \prod_k \widehat C_{d_k,w_k}\,,
\label{9}
\end{align}
and an unknown proportionality constant of order unity. Here $\widehat C_{d,w}$ and $\widehat C_{d_k,w_k}$ are coefficients from either $\mathcal{L}^{(d\le 4)}$ or $\mathcal{L}^{(d)}$, and there is also a sum over all contributions of the form Eq.~(\ref{9}) that  contribute. Standard dimensional analysis gives
\begin{align}
d -4 & = \sum_k (d_k-4)\,,
\end{align}
i.e.\ a graph with two insertions of a dimension-six operator produces a dimension-eight operator, a graph with an insertion of a dimension-six and dimension-two operator produces a dimension-four operator, etc.

Writing the $\mathcal{L}^{(d\le 4)}$ terms in Eq.~(\ref{9}) explicitly gives
\begin{align}
\widehat C_{d,w} & \propto  \left( \widehat C_X \right)^{I_X} \left(\widehat C_\psi \right)^{I_\psi}  \left(\widehat C_\phi \right)^{I_\phi}
\left(\widehat C_M \right)^{V_M}  \left(\widehat C_{m^2} \right)^{V_{m^2}}  \left(\widehat C_y \right)^{V_y}
\left(\widehat C_\kappa \right)^{V_\kappa}\left(\widehat C_\lambda \right)^{V_\lambda}
 \prod_k \widehat C_{d_k,w_k}\,,
\label{creln}
\end{align}
where $I_{X,\psi,\phi}$ are the number of internal $X,\psi,\phi$ lines, and $V_i$ are the number of vertices of type $i$. In strongly coupled theories, $\widehat C_i$ are order unity, and all terms are equally important. However, in weakly coupled theories, some $\widehat C_i$ are small, which gives the usual perturbative expansion. $\widehat C_X$ for the photon in QED is $e^2/(16\pi^2)$,  the  QED expansion parameter. In the NDA rescaled Lagrangian, $e^2/(16\pi^2)$ comes from the photon propagators rather than the vertices, because we are using rescaled fields. Using Eq.~(\ref{chat}) gives
\begin{align}
\widehat C_{d,w} & \propto  \left(\frac{g^2}{16\pi^2} \right)^{I_X} 
\left( \frac{M}{\Lambda} \right)^{V_M}  \left( \frac{m^2}{\Lambda^2} \right)^{V_{m^2}}  \left(\frac{y}{4\pi} \right)^{V_y}
\left( \frac{1}{4\pi} \frac{\kappa}{\Lambda} \right)^{V_\kappa}\left(  \frac{\lambda}{16\pi^2}\right)^{V_\lambda} 
 \prod_k \widehat C_{d_k,w_k}\,.
\label{creln2}
\end{align}
This equation is independent of $I_\psi$ and $I_\phi$ since $\widehat C_\psi$ and $\widehat C_\phi=1$, and each internal gauge field gives a suppresion by $g^2/(16\pi^2)$.
The explict $\Lambda$ terms $M/\Lambda$, $m^2/\Lambda^2$ and $\kappa/\Lambda$ are a reflection that $M,m,\kappa \sim \Lambda$,  are natural values. In the SM, $M$ and $\kappa$ vanish, and $m \ll \Lambda$, the scale of new physics, is the hierarchy problem.

The NDA formula in  Eq.~(\ref{creln2}) can be converted to the usual perturbation expansion in terms of the coefficients $C_{d,w}$ of standard dimensional analysis using the relation
\begin{align}
\frac{C_{d,w}}{\Lambda^{d-4}} &= \frac{\widehat C_{d,w}}{f^{2w} \Lambda^{d-4-2w}}
\end{align}
from Eq.~(\ref{ndaw}), and $\Lambda=4\pi f$, to give
\begin{align}
 C_{d,w} 
& \propto  \left(\frac{g^2}{16\pi^2} \right)^{I_X} \!
\left( \frac{M}{\Lambda} \right)^{V_M} \!
 \left( \frac{m^2}{\Lambda^2} \right)^{V_{m^2}} \!
   \left(\frac{y}{4\pi} \right)^{V_y} \!
\left( \frac{1}{4\pi} \frac{\kappa}{\Lambda} \right)^{V_\kappa} \!
\left(  \frac{\lambda}{16\pi^2}\right)^{V_\lambda}  \!
\left(\frac{1}{4\pi}\right)^{-2w} \!
 \prod_k  \left(\frac{1}{4\pi}\right)^{2w_k} \! C_{d_k,w_k}\,.
\label{ans}
\end{align}
Eq.~(\ref{ans}) is the expression one obtains by direct computation using the original Lagrangian Eq.~(\ref{lag}) and  unrescaled operators 
such as those in the upper row of Eq.~(\ref{ops}). We know that with the unrescaled theory, an $L$ loop diagram is of order $1/(16\pi^2)^L$.
Equating this power of $4\pi$ with the powers of $4\pi$ in Eq.~(\ref{ans}) gives our $4\pi$-counting formula
\begin{align}
2L &= 2I_X + V_y + V_\kappa + 2 V_\lambda -2w + 2 \sum_k w_k\,.
\label{pi}
\end{align}

Equation~(\ref{creln2}) says that an amplitude for a diagram with insertions of the $r$ operators
\begin{align}
O_{d_1,w_1}, \ldots,  O_{d_r,w_r}
\end{align}
to produce the final operator $O_{d,w}$ is 
\begin{align}
\propto \left(\frac{g^2}{16\pi^2} \right)^{I_X} 
\left( \frac{M}{\Lambda} \right)^{V_M}  \left( \frac{m^2}{\Lambda^2} \right)^{V_{m^2}}  \left(\frac{y}{4\pi} \right)^{V_y}
\left( \frac{1}{4\pi} \frac{\kappa}{\Lambda} \right)^{V_\kappa}\left(  \frac{\lambda}{16\pi^2}\right)^{V_\lambda} \,.
\label{areln}
\end{align}
Defining the perturbative order $N$ of a term
\begin{align}
 \left(\frac{g^2}{16\pi^2} \right)^{I_X}   \left(\frac{y}{4\pi} \right)^{V_y}
\left( \frac{1}{4\pi} \frac{\kappa}{\Lambda} \right)^{V_\kappa}\left(  \frac{\lambda}{16\pi^2}\right)^{V_\lambda} 
\qquad \hbox{by} \qquad N = I_X +\frac12 V_y + \frac12 V_\kappa + V_\lambda\,,
\label{porder}
\end{align}
we see that the $4\pi$-counting formula Eq.~(\ref{pi}) gives
\begin{empheq}[box=\fbox]{align}
N &= L + w - \sum_k w_k\, \equiv L + \Delta.
\label{wt}
\end{empheq}
This is our main result. The perturbative order $N$ is not the same as the loop order $L$, but is shifted by 
$\Delta=w-\sum_k w_k$, the difference of the NDA-weight of the final operator and the sum of the NDA weights of the inserted operators.

The NDA formula Eq.~(\ref{wt}) applies to amplitudes (finite and infinite pieces) at general kinematic points. The amplitude can have inverse powers of momenta through particle propagators. In spontaneously broken theories, it can also have inverse powers of the coupling constants (and fields) since particle masses can be proportional to $g \vev{\phi}$ and $y \vev{\phi}$, as in the SM. The formula has to be modified in corners of phase-space, such as near threshold, where one can have kinematic enhancements of $1/v$, where the particle velocity $v$ is a dimensionless number not controlled by NDA.

\section{Application to the Standard Model}

The Lagrangian of the SM EFT has the form
\begin{align}
{\cal L}_{\rm EFT} = {\cal L}_{SM} + \sum_{d > 4} {\cal L}^{(d)},
\end{align} 
where ${\cal L}_{SM}$ is the usual SM Lagrangian with terms of mass dimension $d \le 4$, and ${\cal L}^{(d)}$ consists of all  terms of mass dimension $d>4$.   

There are a total of 59 independent dimension-six operators in the operator basis~\cite{Buchmuller:1985jz,Grzadkowski:2010es}, assuming the conservation of baryon number.  These 59 dimension-six operators divide into eight operator classes~\cite{Grzadkowski:2010es}.  
The coupling constant dependence of the one-loop anomalous dimension matrix $\gamma$ is given in Table~\ref{tab:anom}.
\begin{table}
\renewcommand{\arraycolsep}{0.15cm}
\renewcommand{\arraystretch}{1.5}
\begin{align*}
\begin{array}{cc|cccccccc}
&&g^3 X^3 & H^6 & H^4 D^2 & g^2 X^2 H^2 & y \psi^2 H^3 & g y \psi^2 X H & \psi^2 H^2 D & \psi^4 \\
&& 1 & 2 & 3 & 4 & 5 & 6 & 7 & 8\\
\hline
g^3 X^3 & 1 & g^2 & 0 & 0 & 1 & 0 & 0 & 0 & 0 \\
H^6 & 2 & g^6 \lambda & \lambda, g^2,y^2 & g^4,g^2 \lambda,\lambda^2 & g^6, g^4\lambda  & \lambda y^2, y^4 & 0 & \lambda y^2, y^4 & 0 \\
H^4 D^2 & 3 & g^6 & 0 & g^2 ,\lambda,y^2 &  g^4 & y^2 & g^2y^2 & g^2,y^2 & 0 \\
g^2 X^2 H^2 & 4 &  g^4 & 0 & 1 &  g^2,\lambda, y^2 & 0 & y^2 & 1 & 0 \\
y \psi^2 H^3 & 5 & g^6 & 0 & g^2,\lambda,y^2 &  g^4  &g^2, \lambda,y^2 & g^2\lambda ,g^4,g^2y^2 & g^2 , \lambda  , y^2 & \lambda,y^2 \\
g  y \psi^2 X H & 6 & g^4  & 0 & 0 &  g^2 & 1 & g^2,y^2 & 1 & 1 \\
\psi^2 H^2 D & 7 & g^6 & 0 & g^2,y^2 &  g^4 & y^2 & g^2y^2 & g^2,\lambda,y^2 & y^2 \\
\psi^4 & 8 & g^6 & 0 & 0 &  0 & 0 & g^2y^2 & g^2,y^2 & g^2,y^2 \\
\end{array}
\end{align*}
\caption{\label{tab:anom} The form of the one-loop anomalous dimension matrix $\gamma$ for the coefficients of dimension-six operators in the operator basis 
of Ref.~\cite{Grzadkowski:2010es} when the operators are rescaled according to NDA, see Eq.~(\ref{3.5}).  
The rows and columns of $\gamma$ are labelled by the eight operator classes defined in Refs.~\cite{Grzadkowski:2010es,Jenkins:2013zja}.  
There are a total of 59 operators~\cite{Buchmuller:1985jz,Grzadkowski:2010es} in the eight classes, and we use the notation of 
Ref.~\cite{Jenkins:2013zja}.  Not all the entries are directly due to one-particle irreducible graphs.  Some of them arise by converting operators to the canonical basis using the equations of motion.  This table combines the two tables in Ref.~\cite{Jenkins:2013zja} and also includes wave function renormalization.}
\end{table}

In the SM EFT, Eq.~(\ref{wt}) can be written in a nicer form. The SM EFT Lagrangian has the symmetry
\begin{align}
H & \to -H, & q & \to q, & l & \to l, & u &\to -u, & d & \to -d, & e & \to -e,
\end{align}
where $q,l$ are left-handed quark and lepton doublet fields, and $u,d,e$ are right-handed singlet fields, and the symmetry
\begin{align}
H & \to -H, &  y & \to -y,
\end{align}
where $y$ are the Yukawa couplings. Operators odd under $H \to -H$ can be written instead using $y H$, where $y$ is a formal parameter of order the Yukawa coupling, and is included in the NDA formula Eq.~(\ref{nda}) as a factor of ${y H}/{\Lambda}$
which occurs at most at linear order. Thus the general form of a term in the SM EFT Lagrangian is
\begin{align}
f^2 \Lambda^2 \left(\frac{H}{f}\right)^A \left(\frac{y H}{\Lambda}\right)^{A^\prime} \left(\frac{\psi}{f \sqrt \Lambda}\right)^B
\left( \frac{g X}{\Lambda^2} \right)^C
 \left(\frac{D}{\Lambda}\right)^D
\label{5.2}
\end{align}
with $A^\prime=0,1$.

Operators with $A^\prime=1$ have their NDA-weight shifted by $-1/2$, and coefficient modified by
\begin{align}
\widehat C_{d,w} & \to \frac{f y}{\Lambda}\ \widehat C_{d,w-1/2}
\label{shift}
\end{align}
so that Eq.~(\ref{wt}) remains valid. The advantage is that the $V_y$ in Eq.~(\ref{porder}) is now an even integer, so we obtain instead 
\begin{align}
 \left(\frac{g^2}{16\pi^2} \right)^{n_g}   \left(\frac{y^2}{16\pi^2} \right)^{n_y} \left(  \frac{\lambda}{16\pi^2}\right)^{n_\lambda} ,
 \qquad N = n_g + n_y +  n_\lambda\,,
\label{porder1}
\end{align}
for the perturbative order, since there are no $\kappa$-vertices in the SM. With this convention, the eight operator classes for dimension-six operators in the SM are
\begin{align}
\frac{f^2}{\Lambda^4}\ g^3 X^3,\ \frac{\Lambda^2}{f^4}\ H^6,\ \frac{1}{f^2}\ H^4 D^2,\ \frac{1}{\Lambda^2}\ g^2 X^2 H^2,\
\frac{1}{f^2}\ y\psi^2 H^3,\ \frac{1}{\Lambda^2}\ y\psi^2 g X H,\ \frac{1}{f^2} \psi^2 H^2 D,\ \frac{1}{f^2} \psi^4
\label{3.5}
\end{align}
with NDA-weights $w_i=\left\{-1,2,1,0,1,0,1,1\right\}$ for $i=1,\ldots,8$, which are all integers. This is the normalization convention of Ref.~\cite{Jenkins:2013zja}. We can similarly define weights for the SM  Lagrangian terms in ${\cal L}^{d \le 4}$.  $H^4$ has $w=1$,  and $H^2$ has $w=0$.  The Yukawa operator  $\psi^2 H$ has $w=1/2$ and  can be changed to $y \psi^2 H$ with $w=0$, as in Eq.~(\ref{shift}). The inverse gauge coupling $1/g^2$ multiplies the $w=-1$ gauge kinetic term.

The anomalous dimension matrix  elements
$\gamma_{ij}$, defined by the renormalization group equation for the dimension-six coefficients
\begin{align}
\mu \frac{\rd \widehat C_i}{\rd \mu} &= \gamma_{ij}\, \widehat C_j\,,
\end{align}
have perturbative weight $N_{ij}$
\begin{align}
N_{ij}=L+w_i-w_j \equiv L + \Delta_{ij}\,,
\label{nda6}
\end{align}
from $L$-loop diagrams, which is a special case of Eq.~(\ref{ndaw}).

Consider the one-loop anomalous dimension matrix with the rows and columns reordered in decreasing order of NDA weight $w_i$. The $w=2$ operator is $H^6$ with $i \in \left\{2\right\}$, $w=1$ operators are $H^4 D^2$, $y \psi^2 H^3$, $\psi^2 H^2 D$, $\psi^4$ with $i \in \left\{3,5,7,8\right\}$, $w=0$ operators are $g^2 X^2 H^2$ and $g y \psi^2 XH$ with $i \in \left\{4,6\right\}$, and the $w=-1$ operators is $g^3 X^3$  with $i \in \left\{1\right\}$.
The one-loop anomalous dimension matrix entries have perturbative order $N$ given by
\begin{align}
\begin{blockarray}{c|c|rccccc}
w & \hbox{operators} & & \left\{2\right\} &  \left\{3,5,7,8\right\} & \left\{4,6\right\} & \left\{1\right\} \\[5pt]
\cline{1-2} & \\
\begin{block}{r|c|r(cccc)c}
   2 \phantom{a} &   H^6 &   \left\{2\right\}\phantom{a} & 1 &  2  & 3 &  4 \\[5pt]
   1 \phantom{a} &    H^4 D^2,\ y \psi^2 H^3,\ \psi^2 H^2 D,\ \psi^4  &   \phantom{a} \left\{3,5,7,8\right\}\phantom{a} &  0  &  1 & 2 & 3 \\[5pt]
    0 \phantom{a} &    g^2 X^2 H^2,\ g y \psi^2 XH &    \left\{4,6\right\}\phantom{a} & -1 & 0 & 1 & 2 \\[5pt]
    -1\phantom{a}  &  g^3 X^3 &    \left\{1\right\}\phantom{a} & -2 & -1  & 0 & 1  \\[5pt]
\end{block}
\end{blockarray}
\label{28}
\end{align}
where the rows and columns are labelled by the operator classes, reordered according to decreasing NDA weight.

The anomalous dimensions in Table~\ref{tab:anom} include terms obtained using the equations of motion. Since equations of motion are obtained by varying the action which obeys NDA, Eqs.~(\ref{nda},\ref{28}) are still valid. For example, the penguin operator $O_P=\overline \psi \gamma^\nu \psi \, gD^\mu F_{\mu \nu}$ is converted by the equations of motion to four-fermion operators $O_q=g^2\, \overline \psi \gamma^\mu \psi\, \overline \psi \gamma_\mu \psi$. $O_P$ is a $w=0$ operator, whereas $O_q$ is a $w=1$ operator, so equations of motion can shift $w$. However, $O_q$ is multiplied by $g^2$, so the value of $N$ is also shifted, and Eq.~(\ref{nda6}) remains valid.

Amplitudes can have inverse powers of coupling, via inverse powers of particle masses, but anomalous dimension diagrams can only depend on non-negative powers of couplings, provided no couplings are included in the operators. This observation implies that the entries in Eq.~(\ref{28}) with negative perturbative order vanish. Thus, Eq.~(\ref{28}) explains the pattern of couplings found in Table~\ref{tab:anom}.  Even though the anomalous dimensions in Table~\ref{tab:anom} all come from one-loop diagrams, the individual entries range from $N=0$ to $N=4$, i.e.\ they range from tree order to 4-loop order, with the pattern given by Eq.~(\ref{28}). The formula also applies to the running of SM coefficients  
due to dimension-six operators, as given in Ref.~\cite{Jenkins:2013zja}.

\begin{figure}
\centering
\begin{tikzpicture}

\draw (0,0) circle (1);

\filldraw (-0.1,0.9) rectangle (0.1,1.1);

\filldraw (210:1) circle (0.075);
\filldraw (330:1) circle (0.075);

\draw (90:1) -- +(60:1.5) ;
\draw (90:1) -- +(120:1.5) ;

\draw[dashed]  (210:1) -- +(210:1.5);

\draw[decorate,decoration=snake] (330:1) -- +(330:1.5);

\end{tikzpicture}
\caption{\label{fig:68} Diagram contributing to the $\psi^2 X H- \psi^4$ anomalous dimension $\gamma_{68}$ given in Ref.~\cite{Jenkins:2013zja}.
The solid square is a $\psi^4$ vertex from $\lsix$ and the dots are gauge and Yukawa vertices from $\mathcal{L}_{SM}$. }
\end{figure}
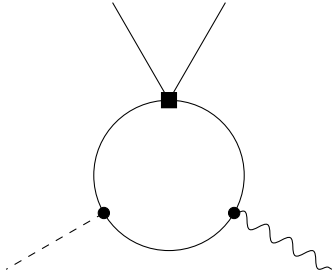

As an illustrative example, Eq.~(\ref{28}) shows that $\gamma_{44}$ computed in Ref.~\cite{Grojean:2013kd} has $N=1$,  and
from Table~\ref{tab:anom}, we see that it has terms of order $g^2,\lambda,y^2$. Similarly
$\gamma_{23}$ has  $N=2$, with terms of order $g^4,g^2\lambda,\lambda^2$;  $\gamma_{24}$ has $N=3$ with terms of order $g^6,g^4\lambda$;
and $\gamma_{21}$ has $N=4$ with terms of order $g^6 \lambda$. There are $N=0$ terms which do not vanish, such as $\gamma_{68}$  from the graph in Fig.~\ref{fig:68} (see Ref.~\cite{Jenkins:2013zja} for more details).

All entries with $N \ge 0$ in Eq.~(\ref{28}) need not vanish. Which entries vanish depends on the  theory under consideration, and its symmetry structure.
For example, $\psi^4$  transforms under flavor symmetries, and only the flavor-singlet $\psi^4$ operators can contribute to the $H^6-\psi^4$ anomalous dimension. In addition, this mixing is forbidden at one-loop for the simple reason that one cannot eliminate four $\psi$ fields by a one-loop diagram.
The mixing, however, will be present at higher loops. As $L$ increases, so do all the numbers in Eq.~(\ref{28}), and eventually, all of them are non-negative so that all the operators can mix at high enough loop-order, if the mixing is not forbidden by  symmetries.

\section{Conclusions}

In conclusion, we have developed a powerful formula based on NDA that constrains the coupling constant dependence
of scattering amplitudes and anomalous dimensions in an arbitrary effective field theory. 
We have used the  SM dimension-six operator running as an illustrative example, but the result is general, and holds for arbitrary EFTs. HQET running of $1/m^2$ operators, including non-linear terms from two insertions of $1/m$ operators, was computed in Ref.~\cite{Bauer:1997gs}, and satisfies the constraint Eq.~(\ref{nda6}).

\acknowledgments

This work was supported in part by DOE grant DE-SC0009919.

\bibliographystyle{JHEP}
\bibliography{nda}

\providecommand{\href}[2]{#2}\begingroup\raggedright\begin{thebibliography}{1}

\bibitem{Grojean:2013kd}
C.~Grojean, E.~E. Jenkins, A.~V. Manohar, and M.~Trott, {\it {Renormalization
  Group Scaling of Higgs Operators and $h \to \gamma \gamma$ Decay }},
  \href{http://xxx.lanl.gov/abs/1301.2588}{{\tt arXiv:1301.2588}}.

\bibitem{Jenkins:2013zja}
E.~E. Jenkins, A.~V. Manohar, and M.~Trott, {\it {Renormalization Group
  Evolution of the Standard Model Dimension Six Operators I: Formalism and
  {$\lambda$} Dependence}},  \href{http://xxx.lanl.gov/abs/1308.2627}{{\tt
  arXiv:1308.2627}}.

\bibitem{Manohar:1983md}
A.~Manohar and H.~Georgi, {\it {Chiral Quarks and the Nonrelativistic Quark
  Model}},  {\em Nucl.Phys.} {\bf B234} (1984) 189.

\bibitem{Buchmuller:1985jz}
W.~Buchmuller and D.~Wyler, {\it {Effective Lagrangian Analysis of New
  Interactions and Flavor Conservation}},  {\em Nucl.Phys.} {\bf B268} (1986)
  621.

\bibitem{Grzadkowski:2010es}
B.~Grzadkowski, M.~Iskrzynski, M.~Misiak, and J.~Rosiek, {\it {Dimension-Six
  Terms in the Standard Model Lagrangian}},  {\em JHEP} {\bf 1010} (2010) 085,
  [\href{http://xxx.lanl.gov/abs/1008.4884}{{\tt arXiv:1008.4884}}].

\bibitem{Bauer:1997gs}
C.~W. Bauer and A.~V. Manohar, {\it {Renormalization group scaling of the
  {$1/m^2$} HQET Lagrangian}},  {\em Phys.Rev.} {\bf D57} (1998) 337--343,
  [\href{http://xxx.lanl.gov/abs/hep-ph/9708306}{{\tt hep-ph/9708306}}].

\end{thebibliography}\endgroup

\end{document}